\documentclass[fleqn,trackchanges,twocolumn]{aastex701}

\usepackage{amsmath}
\usepackage{caption}
\begin{document}

\title{Parameter Estimation Limits in Blazars}

\author[0000-0003-1101-8436]{Agniva Roychowdhury}
\affiliation{National Centre for Radio Astrophysics - Tata Institute of Fundamental Research, Pune University Campus, Ganeshkhind, Pune 411007, MH, India}
\email[show]{agniva.physics@gmail.com}  

\begin{abstract}
Parameter degeneracy in blazar spectral energy distributions (SEDs) is known but rarely quantified. This paper introduces a Fisher Information approach to determine theoretical limits to information extraction in the context of one-zone models. By evaluating the total Fisher Information by varying $\delta$, $B$, $p$, $\gamma_{\rm min}$ and $\gamma_{\rm max}$, we find that EC models encode Fisher information $\gtrsim10^4$ times less than that in SSC models, establishing differences in limits of physical information extraction even in the case of perfect sampling. Moreover, the Fisher information in both SSC and EC models exhibit strong fluctuations across the parameter space, but since the magnitudes are orders of magnitude lower in EC, limits of parameter inference are expected to be worse in FSRQ SEDs than BL Lacs. We also find that the Doppler factor $\delta$ carries at least $10^{2-3}$ more Fisher information than that for $p$ and $B$ in both EC and SSC, making $\delta$ the most constrained SED parameter. Applying our Fisher Information motivated framework to real flaring SEDs of Flat Spectrum Radio Quasars (FSRQs) CTA 102 and 3C 279, we show that mild variations in $\delta$ and $p$ can appreciably produce the flaring SEDs starting from the quiescent model, while in two other flares in 3C 279 simple geometric and spectral considerations cannot reproduce the flares, reducing the efficacy of one-zone models. We propose that time-resolved SED models are indispensable to constraining physical parameters in EC-dominated blazars. 
\end{abstract}

\section{Introduction}

Blazars, a class of active galactic nuclei with closely pointed relativistic jets \citep{blandford2019}, can generally be categorized into two broad types, BL Lacertae Objects (BL Lacs) and Flat Spectrum Radio Quasars (FSRQs) on the basis of their spectral energy distribution (SED) \citep[e.g.,][]{fos98, keen20}. High-energy emission is dominated by Synchrotron and Synchrotron Self-Compton (SSC) in the former while it is External Compton (EC), which is the Comptonization of photons external to the jet, in the latter.

While straightforward one-zone modelling (assumption of a compact emitting region filled with relativistic electrons for example) is generally able to capture the basic shape and structure of a blazar spectrum, the SED data suffers from poor sampling making simple one-zone models highly degenerate in parameter space. Further, even in the case of perfect sampling, the degeneracies might still remain, with certain combinations of the parameters producing similar reduced $\chi^2$. In rare cases of anomalous SEDs (see \citealt{royc22} e.g.,), multi-zone models or more complex one-zone models may be required, which introduce additional degeneracies. In Gamma-Ray Burst (GRB) afterglow studies such situations have been explored more thoroughly \citep{garcia24} whereas the same is particularly lacking in the blazar SED modelling field. This paper aims to produce quantifications of the amount of information that can be extracted from blazar SEDs about specific physical parameters for the two classes of blazars, BL Lacs and FSRQs using a Fisher Information Framework (FIM) \citep{fisher35}. The FIM is a useful diagnostic to ascertain the accuracy of estimating parameters from any given dataset, regardless of sampling strength. Although heavily used in estimation of cosmological parameters (\citealt{tegmark97, tegmark97b} and citations thereto), there is no published work on the same for multi-wavelength blazar data. Further, through two examples of famous FSRQs (CTA 102 and 3C 279) we devise particular solutions to attempt degeneracy breaking especially when their quiescent SEDs can be described by one-zone models well.

The paper is structured as follows. Section \ref{sec:fisher} discusses the Fisher Information methodology. Section \ref{sec:resd} shows the results of our treatment for fiducial SSC and EC models, summarizes the limits of breaking parameter degeneracy and shows corresponding applications to real blazars. Section \ref{sec:conc} ends with a summary of the conclusions of this study.

\section{Fisher Information Framework}
\label{sec:fisher}
Given a collection of data $\mathbf{x}$ the Fisher Information determines the average steepness of the likelihood function around the maximum likelihood point. If $\mathbf{\Theta}=\{\theta_i\}$ represents the parameter space, the Fisher Information Matrix (FIM) \citep{fisher35} is given as :

\begin{equation}
\mathcal{F}_{ij}=-\left\langle\frac{\partial^2 \mathcal{L}}{\partial \theta_i \partial \theta_j}\right\rangle
\end{equation}

where $\mathcal{L}$ is the negative logarithm of the likelihood function $-\ln \mathbf{L}$. The covariance matrix is given as $\mathcal{C}=\mathcal{F}^{-1}$, with the minimum standard deviation on the measured parameter being $\Delta\theta_i\geq(\mathcal{F}^{-1})_{ii}^{1/2}$ (the Cramér-Rao Bound, \citealt{pathak92}), when all other parameters are also derived from the data. The Cramér-Rao Bound quantifies the maximum amount of information one can attain about a parameter regardless of the method or the estimator being used. Therefore, the diagonal elements of the FIM are fundamental measures of the uncertainty on derived parameters.

We deal with synthetic datasets in this paper and we have hence simplified $\mathcal{F}$ for our purpose. The starting model is assumed to be the best-fit for an arbitrary real SED. We perturb the maximum likelihood point through a perturbation fraction $\epsilon=1\%$ to each parameter in order to deduce the local curvature of the likelihood surface for that given parameter. Starting with $\mathbf{L}=-\frac{1}{2}\sum_k \Big[\frac{F_{\rm obs,\nu_k}-F_{\rm model,\nu_k}}{\sigma_k}\Big]^2$, where $F$ is the flux density at a given frequency and $\sigma_k$ is a putative measurement error, after absorbing $\sigma_k$ one can write down the corresponding $\mathcal{F}$ where $\langle F_{\rm model,\nu_i}-F_{\rm obs,\nu_i}\rangle=0$ as 

\begin{equation}
    \mathcal{F}_{\rm ij}=\sum_k \mathcal{R}_i(\nu_k)\mathcal{R}_j(\nu_k)
\end{equation}

where $\mathcal{R}$ is the response function of logarithmic fluxes to logarithmic changes in $\theta_i$, given as $\mathcal{R}_i=\frac{\partial\ln F_{\rm model}}{\partial\ln\theta_i}$, where we took $\epsilon=1\%$ to suppress any higher-order derivatives in the perturbative expansion of $\ln F(\mathbf{\Theta}+\epsilon\mathbf{\Theta})$. The version of the FIM in the previous equation assumes that the parameter space is normally distributed, which may not always be the case. However, since we are probing the response of the logarithmic SED, such effects will be minimized.

Further, to ascertain the \textit{total} Fisher information for different underlying physical states, we vary the more important input parameters of the model defined in Table \ref{tab:fiducial_params} simultaneously for SSC and EC and then compute the corresponding determinant of the Fisher Information Matrix : $\det{\mathcal{F}}$ for each case. This represents a \textit{total} information that includes the $\mathcal{F}$ for all model parameters, and is akin to the information maximization paradigm used in ``D-Optimal Designs" \citep{agu95}.

\section{Results}
\label{sec:resd}

\begin{figure*}[ht!]
    \centering
    \hbox{
    \includegraphics[width=0.5\linewidth]{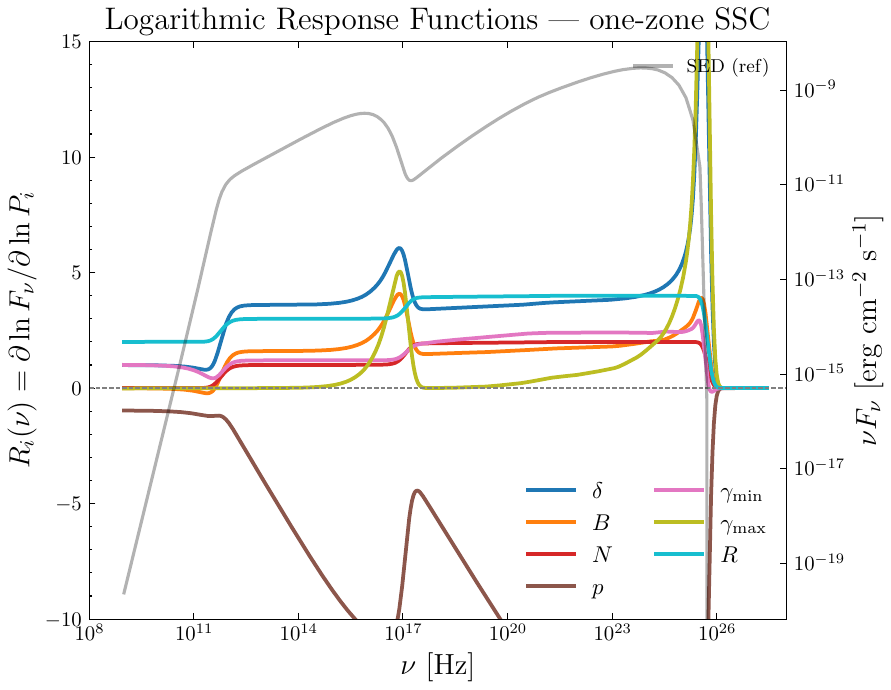}
    \includegraphics[width=0.5\linewidth]{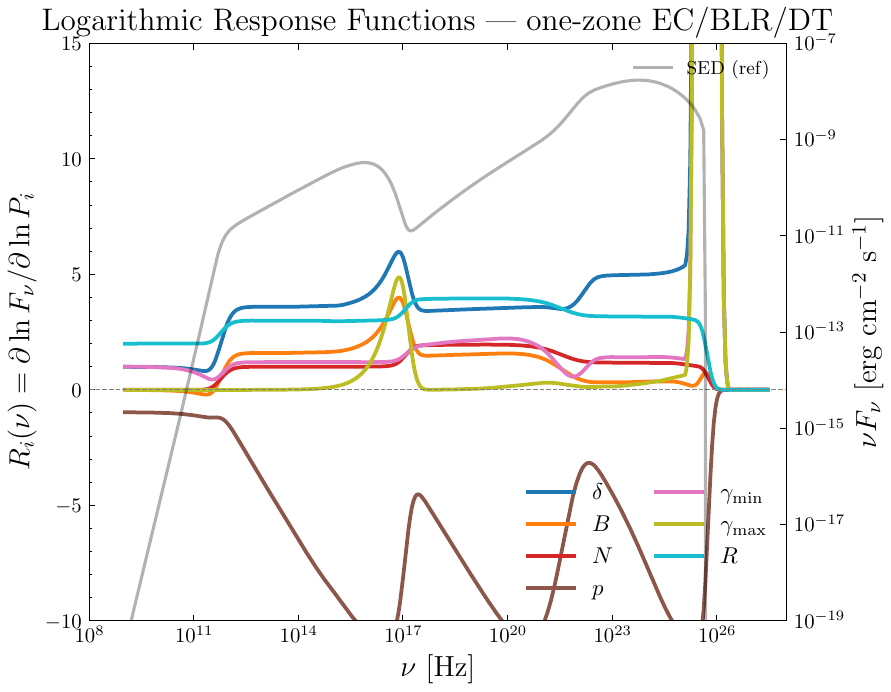}
    }
    \caption{Response functions $\mathcal{R}$ plotted for all relevant physical parameters for the SSC (left) and EC models (right panel). The behaviour of $\delta$ and $p$ are similar in both cases, where they are very sensitive to sharp changes in the spectra. For $\delta$, it is more dominant at EC due to stronger beaming. The magnetic field and electronic energies depict the synchrotron-SSC peak and the cut-off frequencies respectively and their response functions hence show sharp behaviour around those frequencies. $N$ and $R$ control SED normalization and hence do not show sudden changes due to change in model flux across frequency.}
    \label{fig:resp}
\end{figure*}

Figure \ref{fig:resp} shows the response functions $\mathcal{R}$ of the log flux as a function of frequency for a one-zone SSC and a EC-BLR/DT model simulated using \texttt{JetSet} \citep{jetset20}, whose parameters have been detailed in Table \ref{tab:fiducial_params}. Both these models were made to share all parameters common to them, and only the response functions of those have been plotted. Extra EC parameters are not jet parameters per se and hence have been ignored here to enforce simplicity. The left and right panels show the SSC and EC cases respectively. Plotted in gray is the simulated SED for both the models. The $\mathcal{R}$ for the physical parameters have been plotted with different colors as given in the legend of the figure.

As expected, the response functions wildly vary at the cut-off points for the low and high energy components. All parameters except $N$ and $R$ show amplitude variations throughout the entire frequency range, which is expected since they are \textit{normalization} parameters which do not affect the SED shape. One notes that $\delta$ and $\gamma_{\rm max}$ diverge at the high-energy cut-offs for both SSC and EC, implying the high-energy SED is extremely sensitive to these parameters. 

\begin{table*}[ht]
\centering
\caption{Fiducial Parameters for the SSC and EC/BLR Fisher Information Models}
\label{tab:fiducial_params}
\begin{tabular}{lccc}
\hline\hline
Parameter & Symbol & SSC Model & EC/BLR Model \\
\hline
\multicolumn{4}{c}{\textit{Varied Parameters (Fisher Grid)}} \\
Doppler Factor$^*$ & $\delta$ & $30.0$ & $30.0$ \\
Magnetic Field$^*$ & $B$ (G) & $0.5$ & $0.5$ \\
Spectral Index$^*$ & $p$ & 2.2 & 2.2 \\
Min. Lorentz Factor$^*$ & $\gamma_{\min}$ & 55 & 55 \\
Max. Lorentz Factor$^*$ & $\gamma_{\max}$ & $2.5 \times 10^{4}$ & $2.5 \times 10^{4}$ \\
\hline
\multicolumn{4}{c}{\textit{Fixed Jet \& Electron Parameters}} \\
Redshift & $z$ & 0.540 & 0.540 \\
Blob Radius & $R$ (cm) & $1 \times 10^{16}$ & $1 \times 10^{16}$ \\
Dissipation Distance & $R_H$ (cm) & -- & $2 \times 10^{17}$ \\
Electron Density & $N$ (cm$^{-3}$) & 3700 & 3700 \\
\hline
\multicolumn{4}{c}{\textit{Fixed External Field Parameters (EC Only)}} \\
Disk Luminosity & $L_{\text{disk}}$ (erg s$^{-1}$) & -- & $5.0 \times 10^{45}$ \\
Disk Temperature & $T_{\text{disk}}$ (K) & -- & $3 \times 10^{4}$ \\
Inner BLR Radius & $R_{\text{BLR,in}}$ (cm) & -- & $3 \times 10^{17}$ \\
Outer BLR Radius & $R_{\text{BLR,out}}$ (cm) & -- & $5 \times 10^{17}$ \\
BLR Fraction & $\tau_{\text{BLR}}$ & -- & 0.1 \\
DT Radius & $R_{\text{DT}}$ (cm) & -- & $5 \times 10^{18}$ \\
DT Temperature & $T_{\text{DT}}$ (K) & -- & 1000 \\
DT Fraction & $\tau_{\text{DT}}$ & -- & 0.1 \\
\hline
\end{tabular}
\par
\vspace{0.1cm}
\raggedright
\footnotesize{\textbf{Note.} --- Parameters marked with an asterisk ($^*$) represent the fiducial coordinate around which the $5^5$ parameter grid was constructed to evaluate the Fisher Information Structure. All other parameters were held fixed.}
\end{table*}

\begin{figure*}[ht!]
    \centering
    \vbox{
    \includegraphics[width=\linewidth]{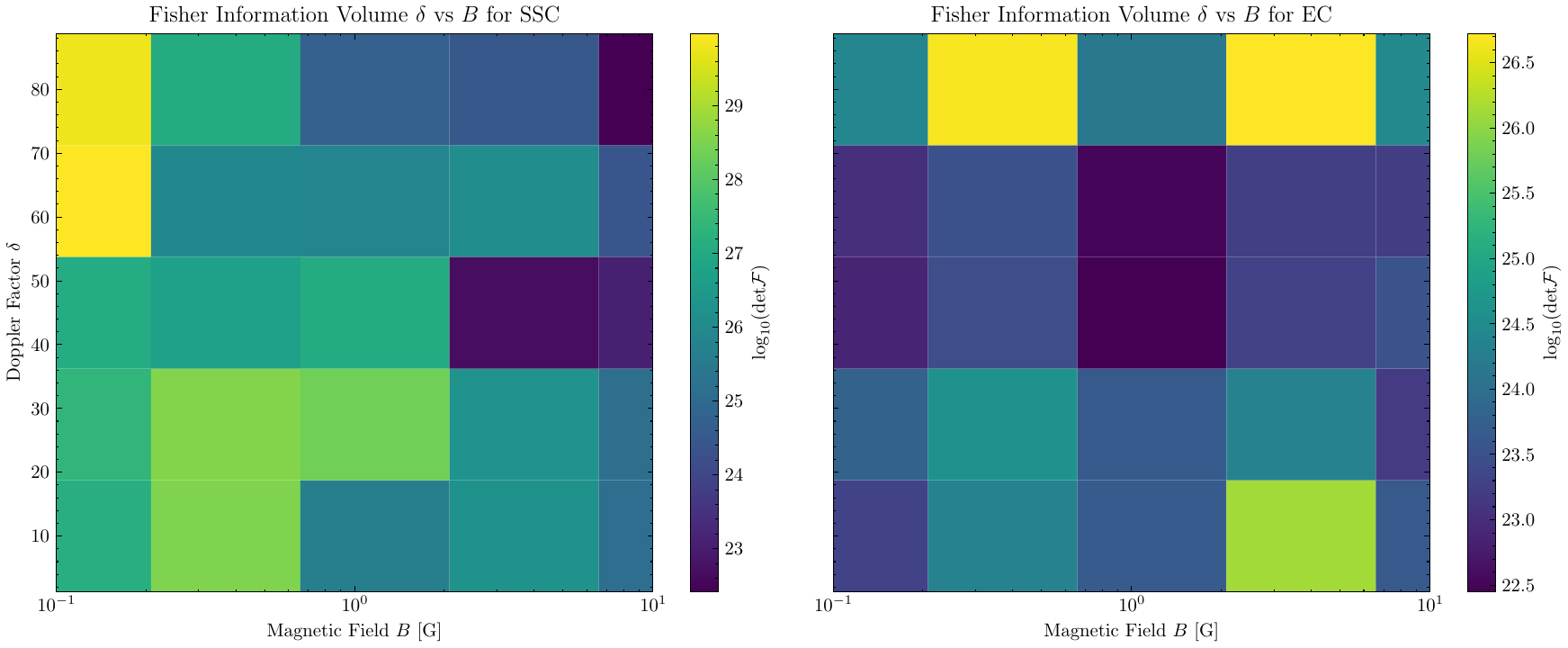}
    \includegraphics[width=\linewidth]{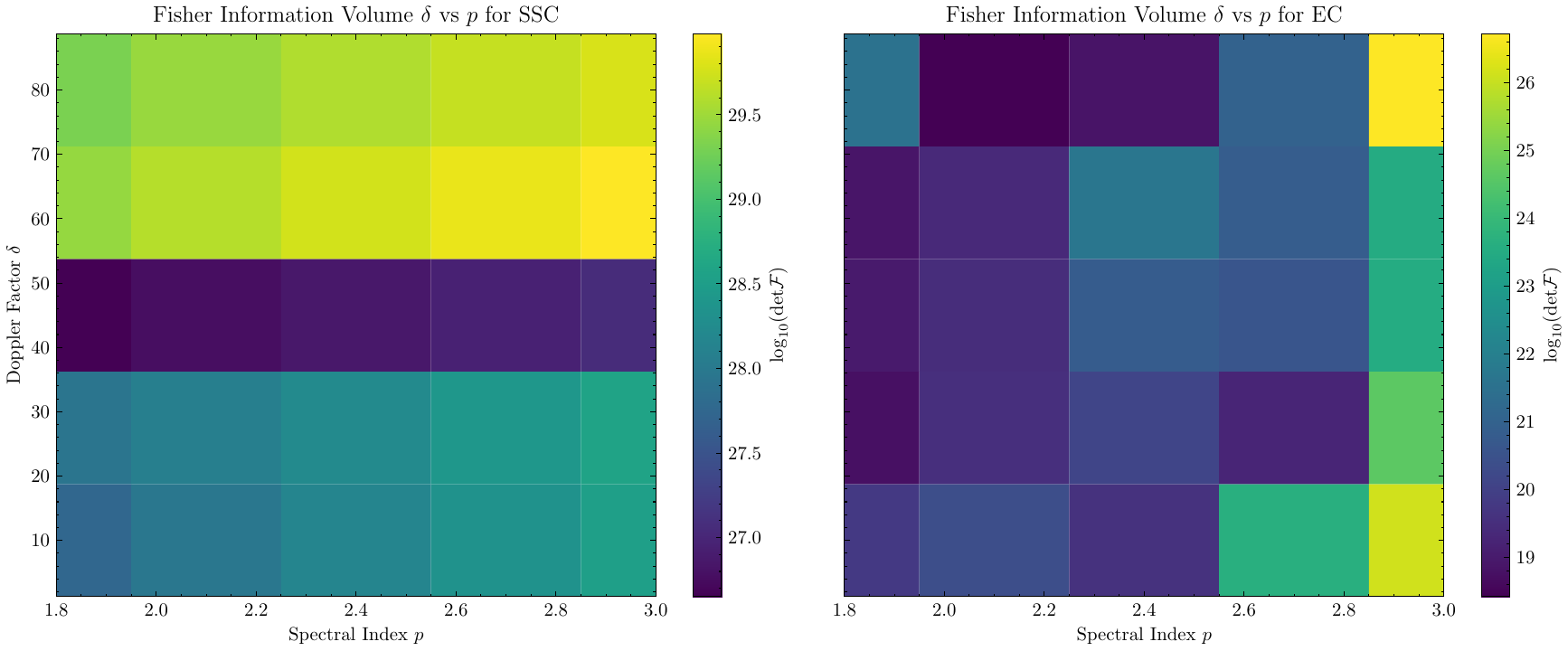}
    }
    \caption{Figure shows the total Fisher information, encoded in $\det\mathcal{F}$ for different parameter sets, marginalized over the maximum values for all the other parameters not displayed in the figure. The Total Fisher Information for $\delta$ v/s $B$ and $p$ (upper and lower panels respectively) for SSC and EC are shown in left and right panels respectively. The Fisher information is orders of magnitude higher $\gtrsim 10^{3-4}$ in SSC models than EC models.}
    \label{fig:delBp_vol}
\end{figure*}

To probe the entire physical space of Fisher Information for the most important SED parameters, namely $\delta$, $B$, $p$, $\gamma_{\rm min}$ and $\gamma_{\rm max}$, we varied these parameters to produce $5^5$ model realizations and therafter computed the five-dimensional Fisher Information Map filled with the determinant of the Fisher Information Matrix that encodes the ``total" Fisher Information. The external photon fields in EC models were kept fixed simply to isolate the \textit{shared} responses between SSC and EC models. Of course, if the photon field is removed, the EC model will behave exactly like the SSC model, and if increased beyond physical limits, the accretion disk emission will overpower the lower-energy emission. Figure \ref{fig:delBp_vol} shows the corresponding $\det{\mathcal{F}}$ as a function of $\delta$ and $B$ (upper panel) or $p$ (lower panel) for SSC (left panel) and EC (right panel) after marginalizing over the maximum of $\gamma_{\rm min}$, $\gamma_{\rm max}$ and either of $B$ or $p$. This is done to exhibit the maximum possible Fisher information. The multiplicative difference in total Fisher information between SSC and EC is massive for both cases of $\delta$ v/s $B$ and $p$, where $\det{\mathcal{F}}$ for the SSC model exceeds that of the EC model by $\sim10^{3-4}$ on an average. Since the total Fisher information is much lower in EC models than SSC, it is natural to conclude that even for a perfectly sampled SED, several different one-zone models could provide a very similar $\chi^2$, because the parameter-averaged curvature of the likelihood surface is much flatter than that for SSC, precluding a partial understanding of the degeneracies themselves. 

\begin{figure*}[ht!]
    \centering
    \includegraphics[width=\linewidth]{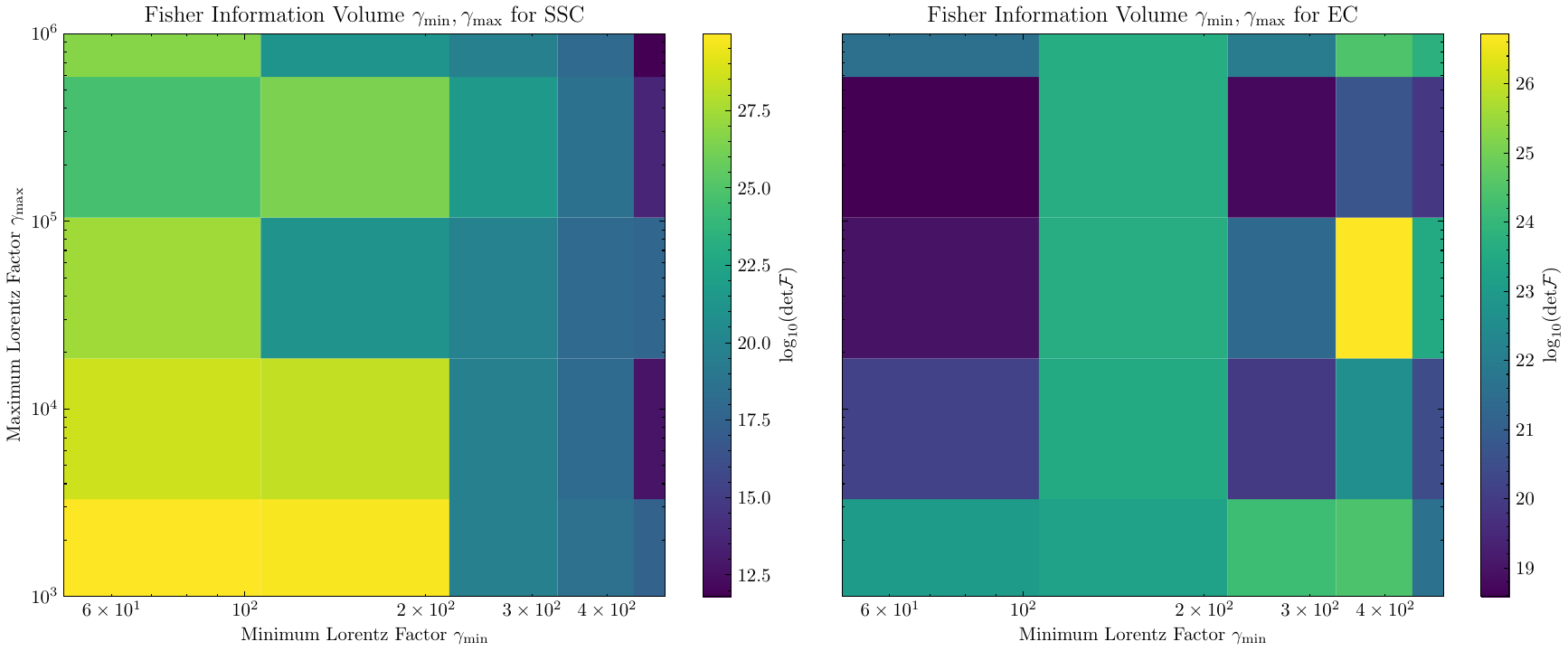}
    \caption{Figure shows the total Fisher information, encoded in $\det\mathcal{F}$ for $\gamma_{\rm min}$ v/s $\gamma_{\rm max}$ for SSC and EC (left and right panels respectively). The Total Fisher Information for $\delta$ v/s $B$ and $p$ (upper and lower panels respectively) for SSC and EC are shown in left and right panels respectively. The average Fisher information in SSC and EC models are very similar, but the lowest values of SSC are orders of magnitude smaller than that in EC. One  notices a drop in the information at larger values of $\gamma_{\rm min}$ perhaps expected from smoothening out of the \textit{frequency gap} between the low and high-energy components.}
    \label{fig:gamma_vol}
\end{figure*}

Figure \ref{fig:gamma_vol} shows $\det{\mathcal{F}}$ for the two cases for $\gamma_{\rm min}$ v/s $\gamma_{\rm max}$. There is no clear pattern in either of the SSC or EC models, but on an average the total Fisher information reaches $\sim 10^4$ times higher values in the SSC case than the EC case. One notices that in both the models, higher values of $\gamma_{\rm min}$ become more difficult to constrain, which is expected since higher minimum energies tend to reduce the \textit{valley width} between the low and high-energy components.

We have included in Appendix \ref{app:A} the plots for both diagonal elements ($\mathcal{F}_{\theta\theta}$) of the FIM and off-diagonal ($\mathcal{F}_{\theta_i\theta_j}$) correlations between different parameters. We find that $\delta$ contains $10^{2-3}$ times more information $B$, for both SSC and EC models, and is rarely degenerate with other parameters, with $\langle\mathcal{F}_{\delta B,\,p}\rangle\ll \langle\mathcal{F}_{\delta \delta}\rangle$, while the absolute correlations are a few orders of magnitude higher in the EC case than in the SSC case. $p$ shows an average Fisher information of $\sim10^{4.5}$ with maximum $10^5$ for the SSC model. The average for the EC model is slightly higher, but instead the values range from $10^{4-8}$. Further, it is evident from the figures in Appendix \ref{app:A} that the SSC self-FIM values for $p$ and $B$ have a more predictable clear pattern, with a monotonic increase in the direction of lower $B$. The corresponding EC cases follow a strong ``checkerboard"-like pattern, suggesting model-fitting becomes more fragile and the maximum likelihood surface changes curvature pseudo-randomly from one parameter space to another. This is in marked contrast to what is observed for SSC models, where the FIM map for both $p$ and $B$ show a very smooth pattern, where the FIM values increase in a particular direction, regardless of the $\delta$ variation. Further, note that the Fisher Matrices were computed using the logarithmic flux response functions, thereby normalizing each parameter to the same scale. The total FIM captures the entirety of the parameter space and we plot it only as a function of the five varying parameters. This is proven in separate plots too, as in Appendix \ref{app:A}, where the magnitudes of the self Fisher information for $p$ are higher than that for $B$, even though the values of $B$ were larger than that of $p$ ($1.8-3.0$ compared against $0.1-10.0$ G). 


\section{Discussion and Conclusion}

\label{sec:conc}

Our analysis of blazar one-zone models using the  Fisher information framework directly tells us that SSC models should be easier to fit and parameters more constrainable than a more complex EC model. This is not unexpected as such since an EC source contains the spectrum of the external photon fields, and the inverse-Compton spectrum of the same, in addition to synchrotron and SSC itself, thereby making fitting more complex since each of these components can interact differently and still produce the same total SED. In an SED, one only observes the \textit{total} flux at each frequency and for an EC model that naturally contains multiple components in addition to an SSC model, even for a perfectly sampled SED, we find that there is a worse upper limit to the amount of physical information one can obtain from sources that are EC-dominated compared to those that are SSC-dominated. Note that in this work the parameters were assumed to be \textit{independent}, which is generally done in SED fitting where parameters are allowed to be varied independently of each other. Physically motivated parameter coupling, which may include the equipartition assumption or a relation between the magnetization and the EED index, will increase the Fisher Information.

\cite{abdo11} is a classic reference that provides population-based details of SED fitting to both BL Lacs and FSRQs. It is relevant to note that they mention more difficulties fitting one-zone models to FSRQs appreciably as opposed to BL Lacs. Note that this is expected simply from more complexity in FSRQ spectra. This complexity works in two ways : first, for a fully sampled perfect FSRQ SED, even if a one-zone model works, we show that the best-fit parameters are highly susceptible to small perturbations, both in the SED and the parameters themselves. A global ``minima" is difficult to measure precisely because a large number of similarly deep minimum wells exist in the system and this issue goes beyond standard MCMC fits that find global extrema. Second, an issue with one-zone model fitting can be multifold. If there is no well-detected minimum for the one-zone model, i.e., it fits poorly through its entire parameter space, a one-zone model may be regarded insufficient. For sparsely sampled SEDs, which is true for nearly all cases, the problem is hence worsened even further.

We note that the observed Fisher information and/or correlations (inverse of the FIM) could be obtained through MCMC simulations regardless. The idea therein is to use imperfectly sampled data following any given prescription, and starting from the original model, find the results/correlations of the final best-fit where the walkers converge. We have tested this for a sample of 64 walkers and 800 steps and our correlation matrix reproduced the inverse FIM within an accuracy of $\lesssim1\%$ (starting from the 3C 279 best-fit model to the quiescent state, Appendix \ref{app:B}). While we have not included this discussion in this paper for the sake of brevity, we note that a MCMC fit is generally time-consuming and the FIM serves as a preemptive approach to SED fitting, highlighting regions of the parameter space where model-fitting could be conclusive or its opposite.

Further, as we showed in Appendix \ref{app:B} using examples of CTA 102 and 3C 279, a mild tuning of the Doppler factor $\delta$ and $p$ can explain most of the flaring in FSRQs, motivated by the presence of large Fisher information (compared to other parameters in FSRQs/EC-models) in both these parameters (Appendix \ref{app:A}) across a wide range in parameter space.

The question therefore is ``what next"? If FSRQs cannot be as accurately modelled (inspite of having an ideal sampling) as BL Lacs as our results dictate, what does it imply for future state-of-the-art telescope observations? To answer this we refer to the results of Appendix \ref{app:B} again. If the blazar has a well-sampled time-dependent SED, it is the only answer to constraining parameters well. If the quiescent SED is well fit (a good $\chi^2$) it is a priori unclear if that indeed forms the best conclusion to the physical parameters, as presented in Figures \ref{fig:delBp_vol} and \ref{fig:gamma_vol}. But if there are multiple flaring SEDs, one would need to determine the minimum number of parameters that need to be changed to explain the flare. If the change can proceed simply, as in some results of Appendix \ref{app:B}, then the parameters have been constrained well. In contrast, if that requires changing parameters that do not have a strong FIM magnitude (for example, anything other than $\delta$ and $p$ in a one-zone model assuming external photon fields are static), and which rather changes the baseline quiescent model altogether, whether or not it is a good fit, we must conclude it was not the same baseline zone that caused the flare. This would further imply a need for multi-zone models, either at the same time or at different times with different underlying configurations. In any case, a modification of parameters to explain flaring SEDs starting from a steady-state SED would require an informed guess developed from estimating the FIM of the models being used.

\section{Acknowledgments}

We acknowledge the support of the Department of Atomic Energy, Government of India, under the project 12-R\&D-TFR-5.02-0700. ARC acknowledges Akram Touil for insightful comments on the manuscript, Dipanjan Mitra for mentioning the idea of deducing dominant modes for blazar flare production and Kaustav Mitra for mentioning Fisher Information in a casual conversation years back.

\appendix

\section{Fisher Information Diagonal and Off-Diagonal Maps Across $\delta,\,B$ Variation}
\label{app:A}

In order to determine the limits to the estimation of parameters in blazar SEDs, we computed the Fisher information matrix using the response functions for the fiducial SED. The resulting $\mathcal{F}$ matrix will display the information limits to the estimation of the each parameter in the model. However, in order to test the stability of the information bounds, we varied $\delta$ and $B$ (for simplicity) in our fiducial parameter space between $10-20$ and $0.1-10$ G respectively and then computed the corresponding $\mathcal{F_{\delta\delta}}$, $\mathcal{F}_{\rm BB}$ and $\mathcal{F}_{\rm pp}$ (diagonal elements) which would throw a clear light on the accuracy of estimating $\delta$ and $B$ across a range of spectral energy distributions.

\begin{figure}
    \centering
    \hbox{
    \includegraphics[width=0.45\linewidth]{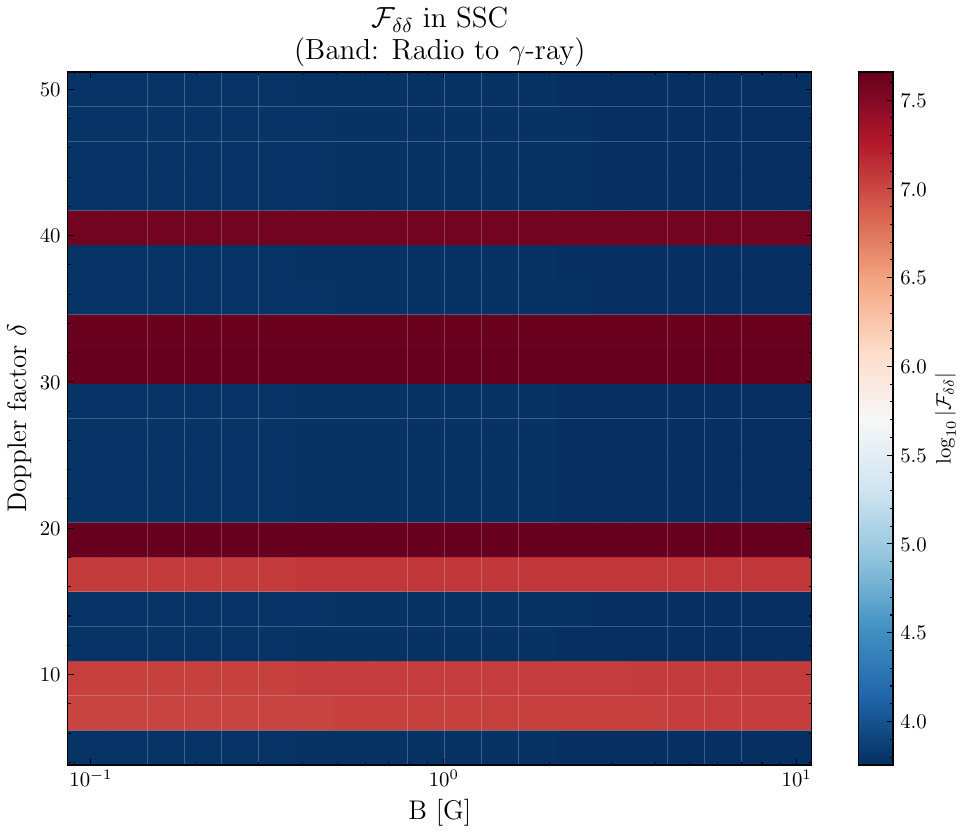}
    \includegraphics[width=0.5\linewidth]{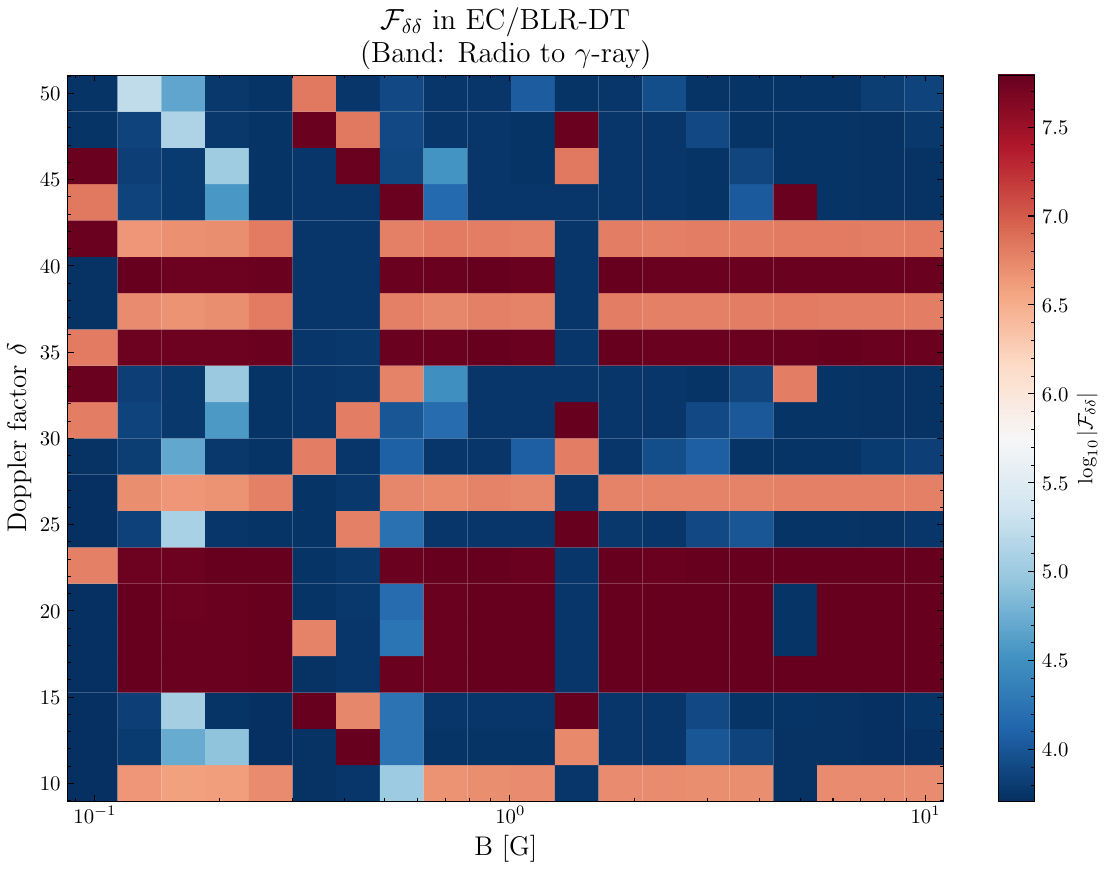}
    }
    \caption{Fisher Information Maps for $\delta$ for the SSC (left) and EC models (right). $\mathcal{F}_{\delta\delta}$ is remarkably high $\sim10^{4-8}$ throughout the two models, suggesting it is the most well-constrained parameter in one-zone SED fits.}
    \label{fig:fim_ssc_ec_del}
\end{figure}

\begin{figure*}[ht!]
    \centering
    \includegraphics[width=0.48\linewidth]{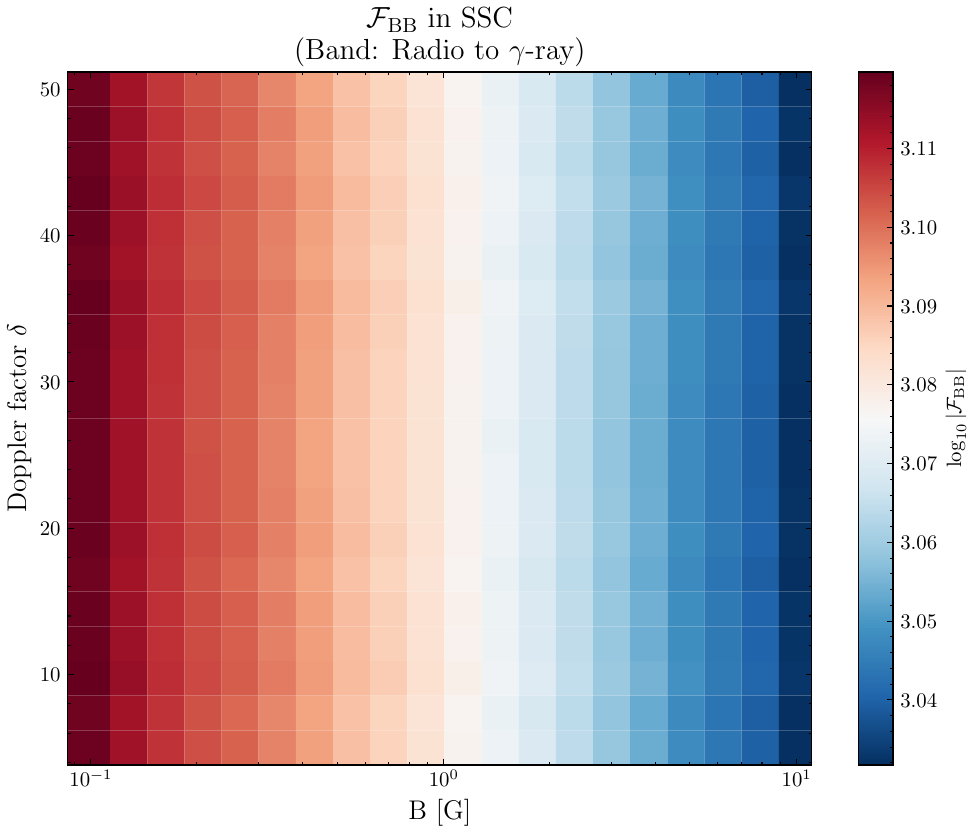}\hfill
    \includegraphics[width=0.48\linewidth]{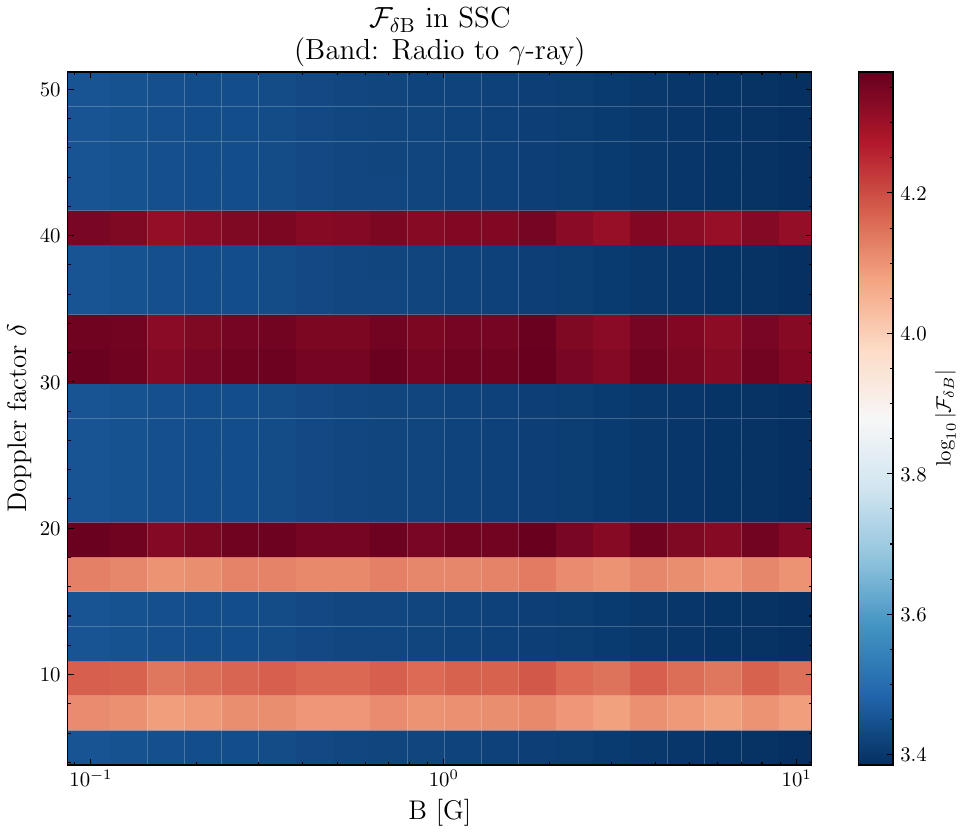}
    
    \vspace{0.2cm} 
    
    \includegraphics[width=0.48\linewidth]{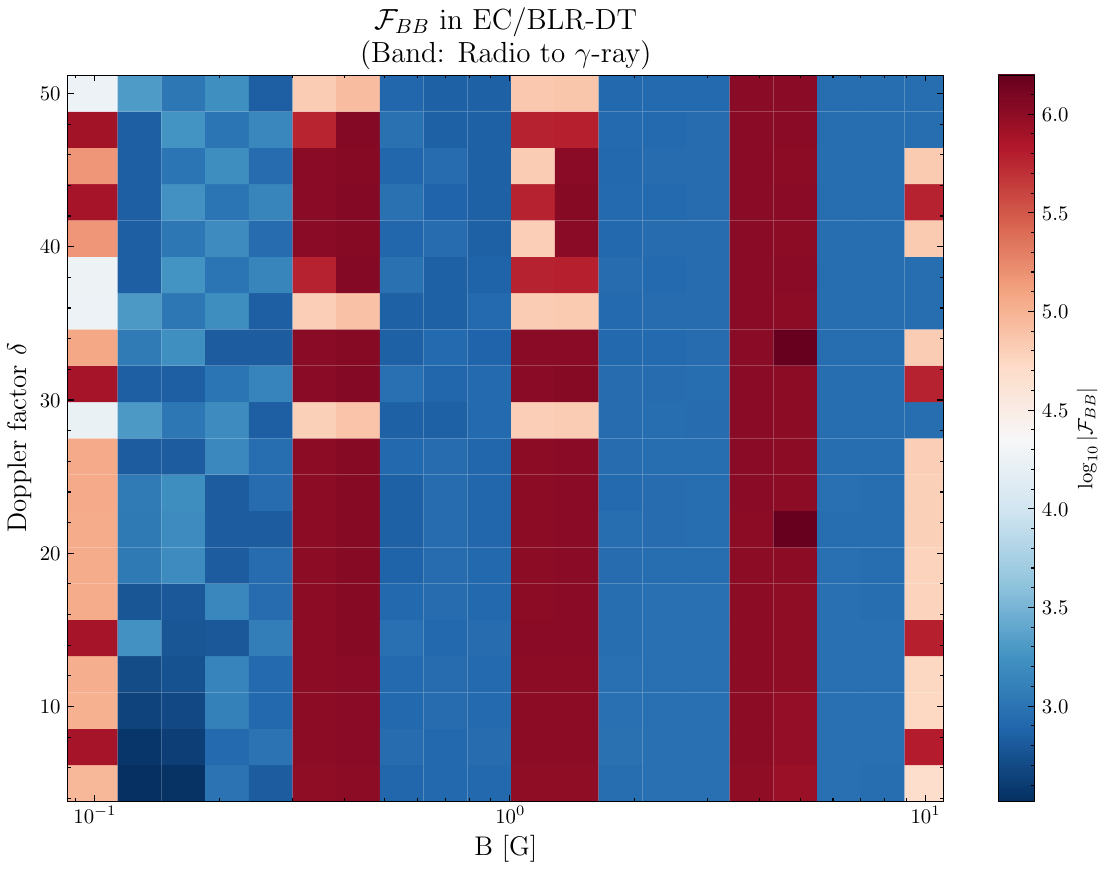}\hfill
    \includegraphics[width=0.48\linewidth]{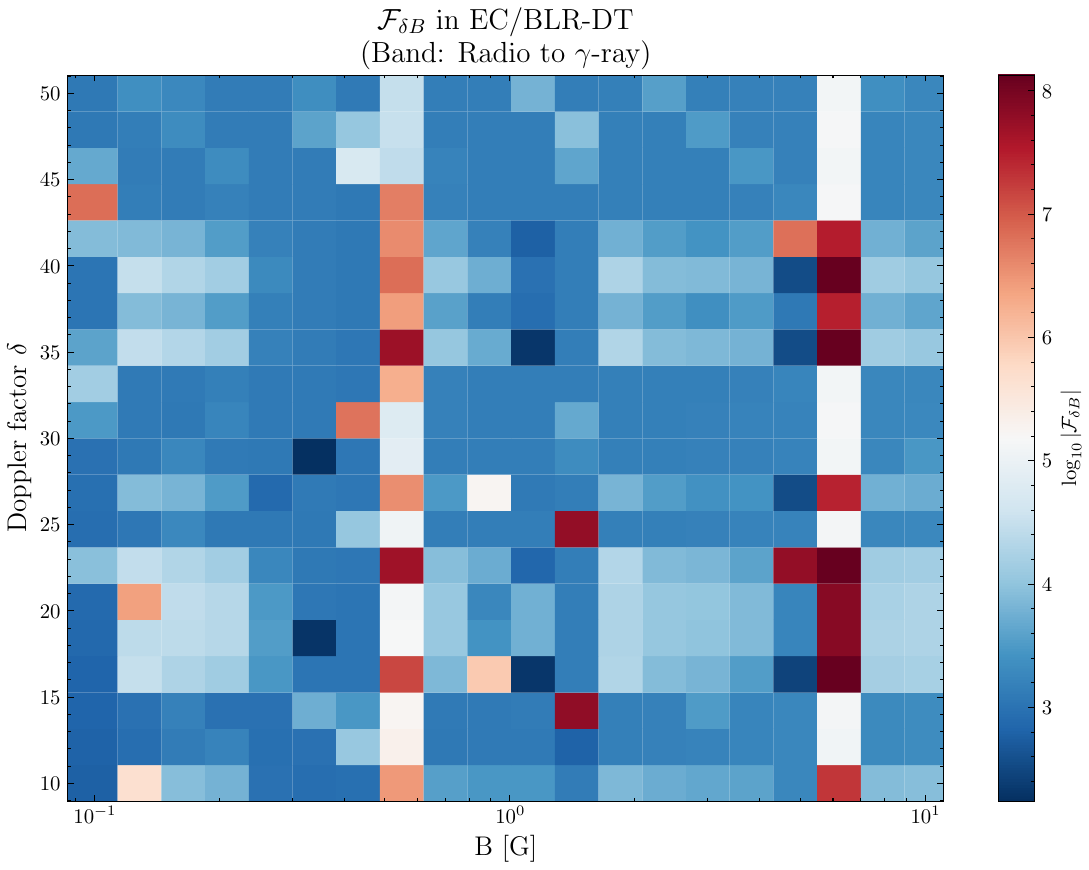}
    
    \caption{Left Panel : Fisher Information Maps for $B$ for the SSC (left) and EC models (right). $\mathcal{F}_{BB}$ is remarkably stable varying only by $10^{0.1}$ throughout the parameter space in SSC. The EC case suggests more fragility/sensitivity to the underlying parameter space, while the values span three orders of magnitude, are larger than that in SSC. Right Panel : The off-diagonal elements of the FIM between $\delta$ and $B$ for each of SSC and EC models. The magnitudes are lower, as expected, than $\sqrt{\mathcal{F}_{\delta B}\mathcal{F}_{\delta\delta}}$. In the SSC case the FIM map shows little variation $\sim 10^1$ and the magnitudes are $\sim10^4$ on average, while in the EC case, the magnitudes are few orders of magnitude larger than SSC, suggesting degeneracies between $\delta$ and $B$ in numerous parts of the parameter space are higher than in the SSC case.}
    \label{fig:fim_ssc_ec_delb}
\end{figure*}

Figure \ref{fig:fim_ssc_ec_del} shows the corresponding plot for $\mathcal{F_{\delta\delta}}$ as a function of parameter movement for the SSC and EC cases. The left and right panels show the SSC and EC cases respectively. $\mathcal{F}_{\delta\delta}$ is remarkably high $\sim10^{4-8}$ throughout the two models. The left panel of Figure \ref{fig:fim_ssc_ec_delb} shows $\mathcal{F}_{\rm BB}$ and $\mathcal{F_{\delta B}}$ for the SSC and EC cases. One first notes that the Fisher information in $\delta$ is almost $\sim 3$ orders of magnitude higher than that in $B$. This implies the curvature of the likelihood surface is $\sim 10^{2-3}$ times more sensitive to changes in the underlying intrinsic $\delta$ than the magnetic field $B$. This difference in magnitude is evident in both SSC and EC, implying for an SED fit, one needs to consider geometric variations, i.e., changes in the viewing angle, more importantly than the magnetic field. The structure of $\mathcal{F}_{\rm BB}$ in SSC shows a clear increase at lower magnetic fields (although only from $\sim$ 3.05 to 3.30). At higher magnetic fields the ratio between the SSC and Synchrotron peak fluxes decrease due to increased ``magnetic compactness" \citep{petrop14}. This leads to a less-pronounced valley, making estimation of parameters that determine the peak frequencies and cut-offs difficult. In contrast, determination of $\delta$ is mostly unaffected by changes in the underlying $\delta$ and $B$. In the EC case, this is even more pronounced. The EC case for $\mathcal{F}_{\rm BB}$ shows a wide range of values spanning three orders of magnitude, with no clear order like in the SSC. There are only few columns where the estimation of $B$ can be made with a high accuracy, and the entire map, across this set of $\delta,B$ values, shows extreme volatility. However, we note that in certain regions of the $\mathcal{F}_{\rm BB}$ map in the EC case the magnitudes are $\sim10^3$ higher than that in the SSC case. It is non-trivial to compute the physical cause behind such an observation since there is no ``clear" pattern unlike the SSC case. But it is nevertheless evident that the $\mathcal{F}_{\rm BB}$ map for the EC case is pretty unstable to underlying changes in the blazar spectrum and is hence difficult to pin using SED model fits to imperfect data. This implies that determination of $B$ requires highly specific physical regimes in blazars which would allow a better determination of $B$. For $\delta$, in contrast, the parameter variations create negligible change in the Fisher information structure. $\mathcal{F}_{\delta\delta}$ maintains high values $\gtrsim10^6$ throughout. This is expected due to the strong dependence of both the SSC and EC fluxes on $\delta^{5/2+p/2}$ and $\delta^{3+p}$ respectively. From the two figures, we come to three conclusions. First, the Doppler factor can be estimated at a much higher accuracy ($\sim 10-10^{3/2}$ times, using the Cramér-Rao Bound, \citealt{tegmark97}) than the magnetic field. Second, estimation of the magnetic field is extremely sensitive to the underlying parameter space. Third, since variations in parameter space cause significant changes in the Fisher information structure of the magnetic field $B$, it is evident that the degeneracy structure, obtained from inverting $\mathcal{F}_{ij}$ will be highly volatile and dependent on the parameters too.

The right panel of Figure \ref{fig:fim_ssc_ec_delb} shows the correlations between $\delta$ and $B$ as a function of physical parameter, and particularly the correlations are larger in the EC case than in SSC.

\begin{figure}
    \centering
    \includegraphics[width=0.48\linewidth]{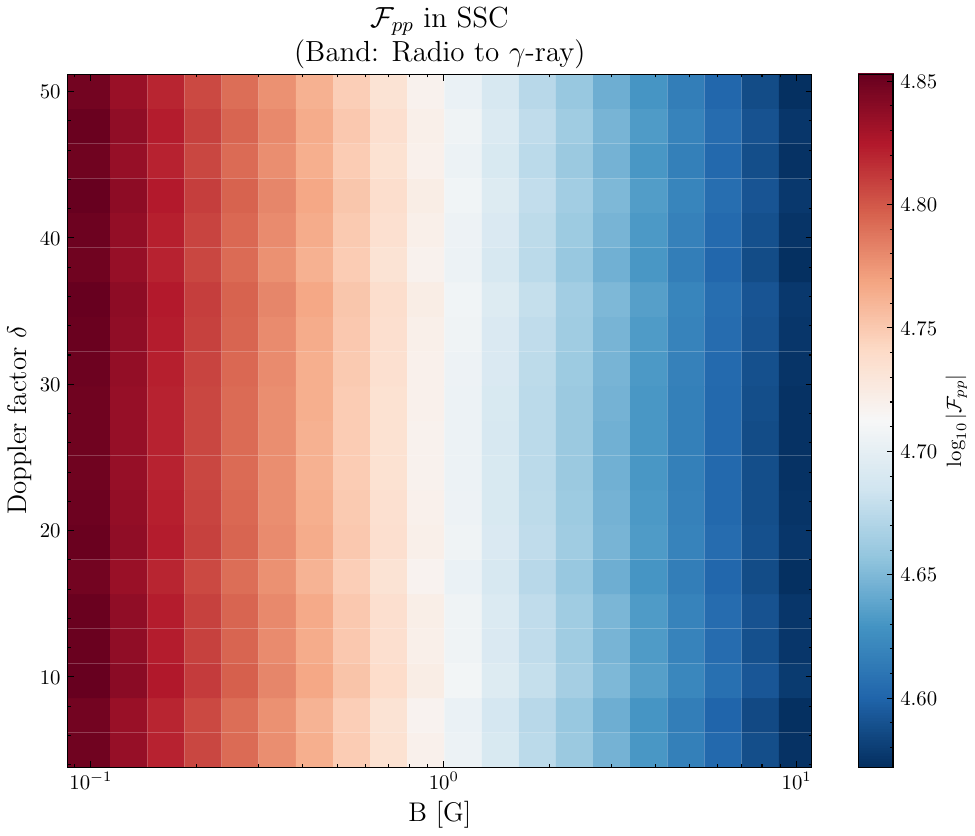}\hfill
    \includegraphics[width=0.48\linewidth]{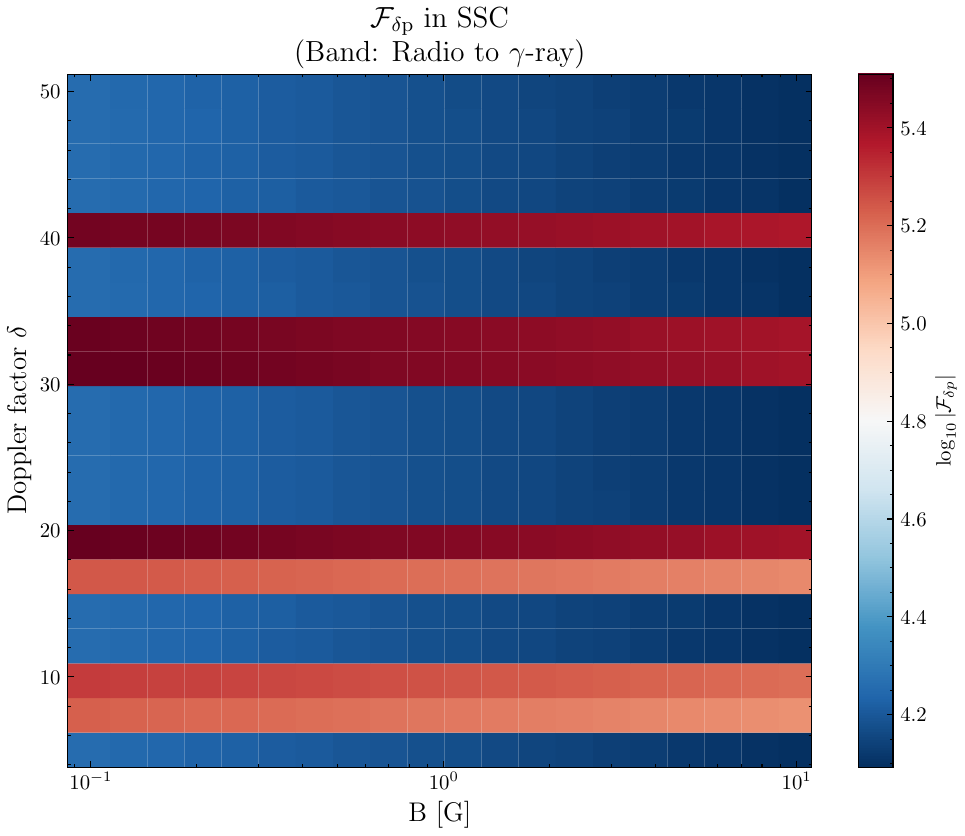}
    
    \vspace{0.2cm} 
    
    \includegraphics[width=0.48\linewidth]{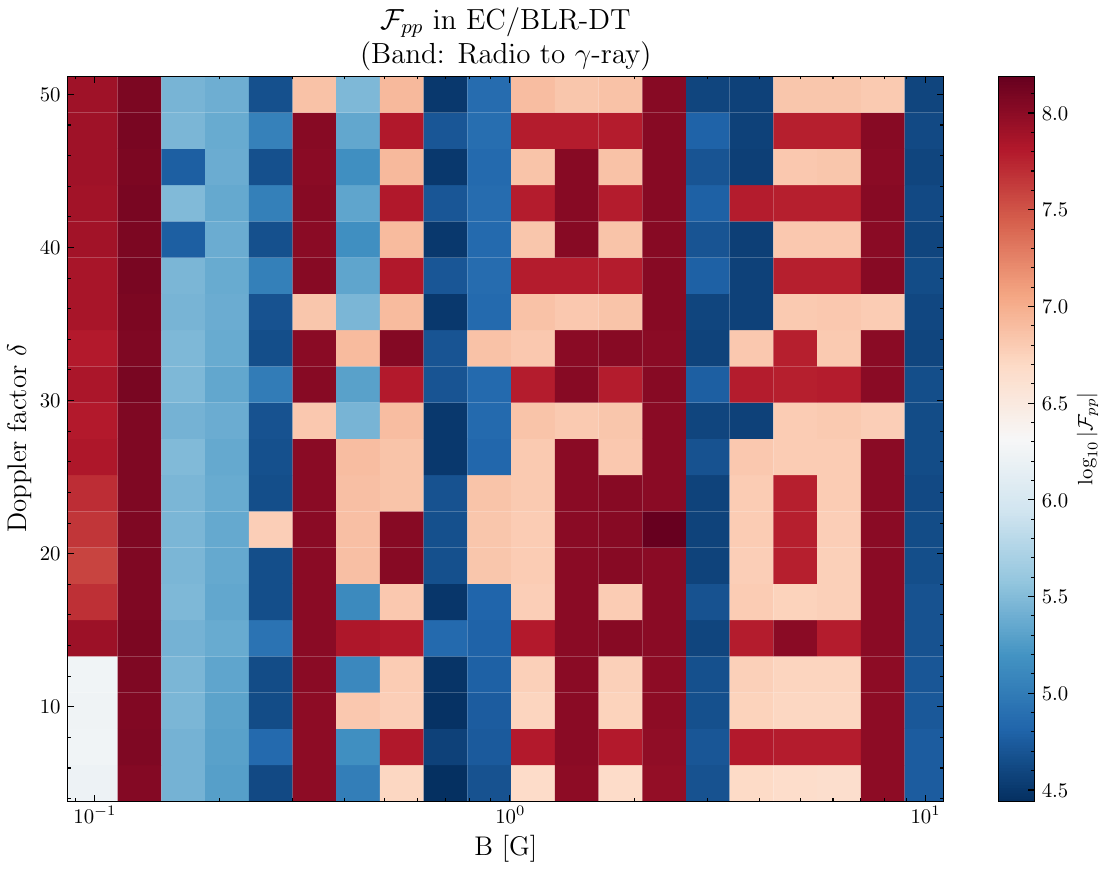}\hfill
    \includegraphics[width=0.48\linewidth]{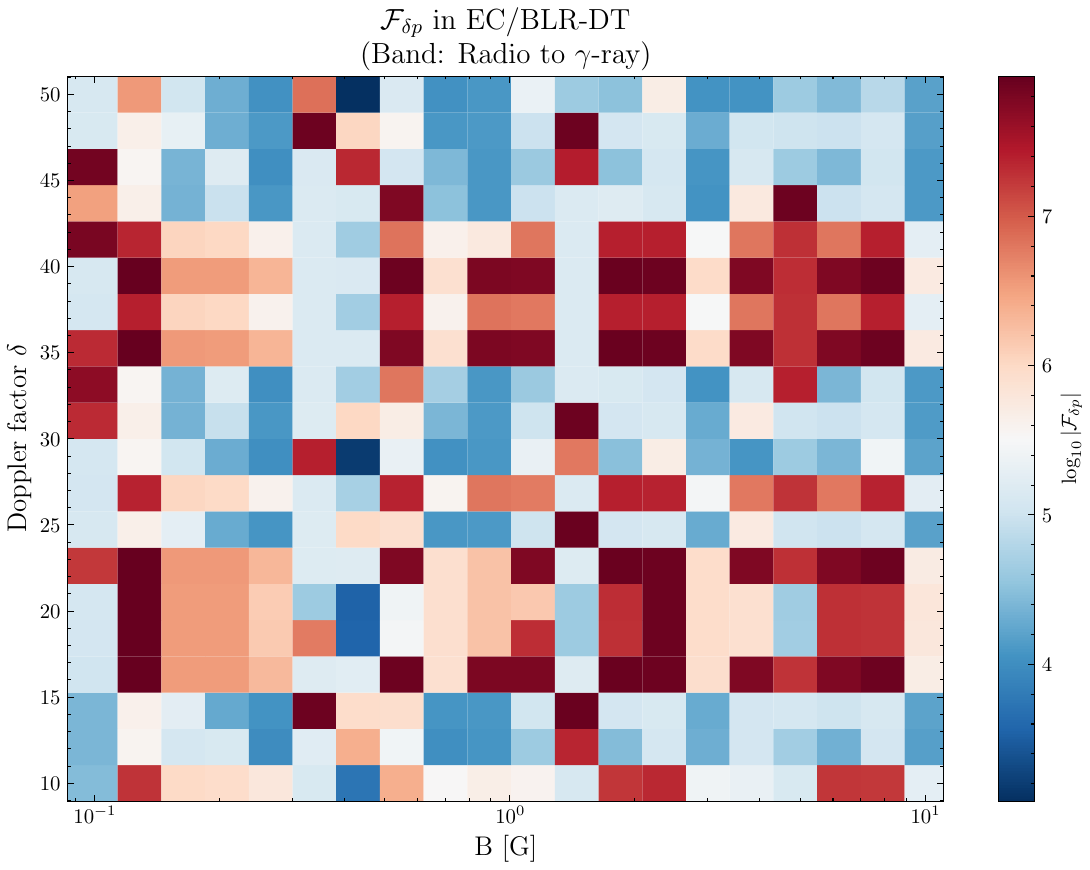}
    
    \caption{$\mathcal{F}_{pp}$ and $\mathcal{F}_{\delta p}$ are displayed in the figure for SSC and EC cases in the left and right panels respectively. Moderately large values of $\mathcal{F}_{pp}$ ($\mathcal{F}_{\rm BB}\lesssim\mathcal{F}_{pp}\lesssim\mathcal{F}_{\delta \delta}$) are observed for both SSC and EC. For correlations between $\delta$ and $p$, the correlations are generally always much less than $\mathcal{F}_{\delta\delta}$ (and lesser in SSC than EC) and sometimes in the order of $10^6$. One notes that in SSC both the degeneracies and diagonal value of the FIM have a certain pattern across the parameter landscape. For EC, in contrast, there is no pattern, implying the curvature of the likelihood surface changes drastically from a small change in the underlying physical model.}
    \label{fig:fim_corr_ssc_ec}
\end{figure}

Figure \ref{fig:fim_corr_ssc_ec} shows the diagonal elements of the Fisher matrix for the EED index $p$, the off-diagonal elements $\mathcal{F}_{\delta p}$ for correlations between the Doppler factor and each of $p$. One observes moderately large values of $\mathcal{F}_{pp}$ ($\mathcal{F}_{\rm BB}\lesssim\mathcal{F}_{pp}\lesssim\mathcal{F}_{\delta \delta}$) for both SSC and EC. For correlations between $\delta$ and $p$, the correlations are generally always much less than $\mathcal{F}_{\delta\delta}$ (and lesser in SSC than EC) and sometimes in the order of $10^6$.  One notes that in SSC, yet again, the diagonal value of the FIM has a certain pattern across the parameter landscape. For EC, both the correlations and the self-FIM magnitudes are larger, implying a possibly better constraining of $p$ than SSC. However, there is no visible pattern in the EC case (scattered), implying the curvature of the likelihood surface changes drastically from a small change in the underlying physical model thereafter making it difficult to pinpoint the best-choice of parameters compared to any given SSC-dominated SED.


\section{Applications to 3C 279 and CTA 102}
\label{app:B}

\begin{figure*}
    \centering
    \hbox{
    \includegraphics[width=0.49\linewidth]{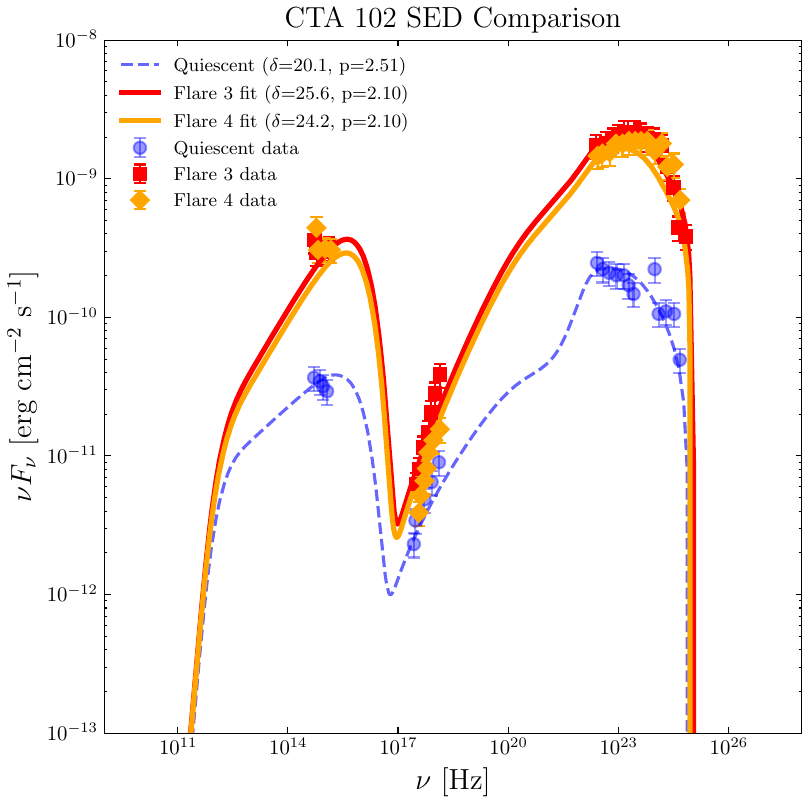}
    \includegraphics[width=0.5\linewidth]{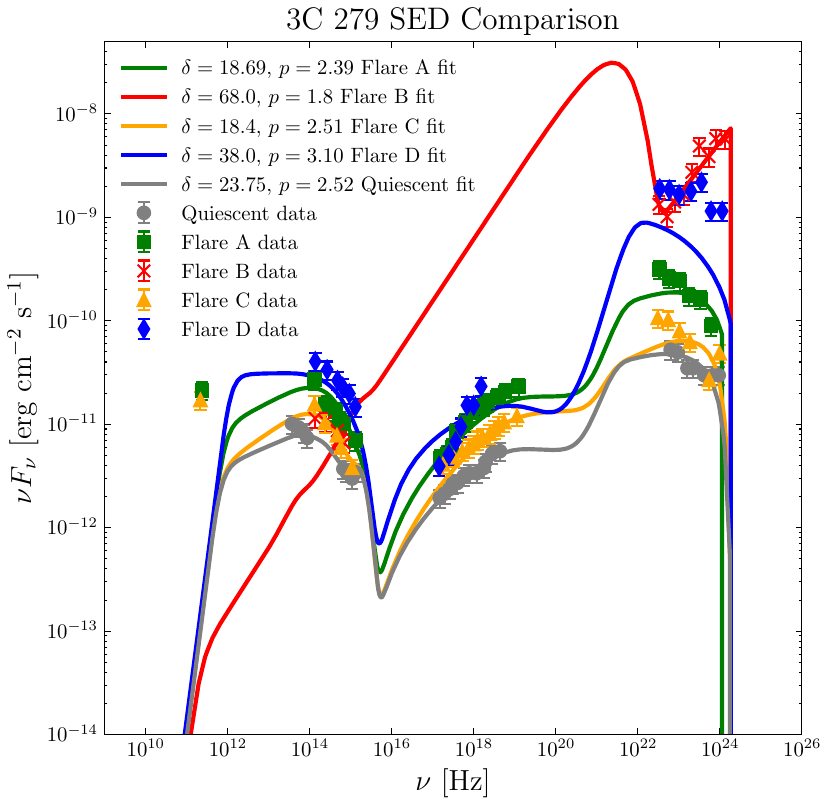}
    }
    \caption{Left: Quiescent and Flaring SEDs of CTA 102 from \cite{prince18}. Starting from the quiescent best-fit model, a slight change in $\delta$ and a 15\% change in $p$ can successfully reproduce the flaring state. Right: For 3C 279, flares A and C could be reproduced by simple geometric (jet-bending) arguments since the spectral index changed only by $\lesssim4\%$. However, flares B and D, and especially B required a very hard EED index and extremely high $\delta$, and flare D cannot be fit well with simple tuning of $\delta$ and $p$, suggesting the need for more complex models.}
    \label{fig:sed}
\end{figure*}

Using the understanding of the Fisher information for different physical parameters in blazar one-zone models, we fit flaring SEDs of two popular FSRQs, CTA 102 and 3C 279, in this subsection. To this end, we use the data published by \cite{prince18} for CTA 102 and \cite{hayashida15} for 3C 279. We note that very recently Doppler alignment and magnetic reconnection were proposed as a possible flaring mechanism and state transitions for the former \citep{royc26}.

We added a flat 20\% error floor to all the data before fitting. Table \ref{tab:quiescent_params} below shows the best-fit parameters for the steady states of both these sources using \texttt{JetSet} \citep{jetset20}. A number of parameters, which are external to the jet, were tuned initially and kept fixed for the model minimizer inside \texttt{JetSet}. Our goal was to determine the minimum amount of tuning to the quiescent parameters in order to reproduce the flaring states of the two source. For this exercise, we used the largest two flares of CTA 102 from \cite{prince18}, which are characterized as ``Flares 3 and 4" in their paper and all 3C 279 flares A, B, C and D from \cite{hayashida15}. We find that the flaring SEDs can be solely represented by changing the Doppler factor $\delta$ and the EED index $p$, as long as the SEDs do not require extreme changes in the spectral index. 

We illustrate this in Figure \ref{fig:sed}, which shows the SEDs of CTA 102 (left) and 3C 279 (right) side by side. For CTA 102, the flares could not be reproduced by a simple $\delta$ tuning, but instead a combined change in $\delta$ and $p$ could easily reproduce both the flare, instead of a $\sim$ 100 time increase in the kinetic luminosity (Table 5 of \citealt{prince18}). The change in $\delta\sim5$ at $\delta=20$ for CTA 102 is easily produced by jet bending of $\sim 0.5^\circ$ if the bulk Lorentz factor is $\Gamma\gtrsim30$. Of course, for much lower $\Gamma$, slight jet bending will not be sufficient to change $\delta$. The EED index required a $\sim15\%$ change to a harder spectrum, possibly indicating shock energization. However, a simultaneous change in $\delta$ and $p$ can also be explained by magnetic reconnection in minijets-in-a-jet, especially if a ``monster" plasmoid orients itself towards us \citep{gian09,royc26,das26}.

For 3C 279 the case is more complex. The figure illustrates the different $(\delta,p)$ pairs that could be used to fit the flaring states. Simple changes in $\delta$ and $p$ recover the flaring states A and C. However, for B and D, and especially B, the required changes are extreme and the fit is poor. An extreme flattening of the spectral index is necessitated to predict the GeV fluxes, as also discussed in \cite{hayashida15}. 


Our conclusions from this analysis are hence two-fold. One, starting from the best-fit physical model of the observed, geometric and spectral index corrections are the easiest to make and provide the strong possibility to explain flares in the simplest way possible. Sub-degree jet bending and mild electron injection through reconnection/shocks are also physically more amenable than a forced increase in the kinetic luminosity, the magnetic field or the emission region size. Further, since $\delta$ is the least \textit{uncertain} parameter in a one-zone model, modifying it to explain time-evolution of SEDs should be the first step. Second, this is only in the context of simple one-zone models. If these fail to describe the shape of the SED, one would need more complex models, like that for Flares B and D in 3C 279.

\begin{table*}[ht]
\centering
\caption{Quiescent SED Best-fit Parameters for 3C 279 and CTA 102.}
\label{tab:quiescent_params}
\begin{tabular}{lccc}
\hline\hline
Parameter & Symbol & 3C 279 (Quiescent) & CTA 102 (Quiescent) \\
\hline
\multicolumn{4}{c}{\textit{Varied Parameters (Fitted)}} \\
Doppler factor & $\delta$ & 11.80 & 20.05 \\
Magnetic field (G) & $B$ & 2.08 & 3.67 \\
Electron density (cm$^{-3}$) & $N$ & 3707.2 & 1853.9 \\
Spectral index & $p$ & 2.64 & 2.51 \\
Min. Lorentz factor & $\gamma_{\min}$ & 54.63 & 89.86 \\
Max. Lorentz factor & $\gamma_{\max}$ & 2496.5 & 7981.5 \\
Blob radius (cm) & $R$ & $1.01 \times 10^{16}$ & $7.26 \times 10^{15}$ \\
Disk Luminosity (erg s$^{-1}$) & $L_{\text{disk}}$ & $5.38 \times 10^{45}$ & $6.71 \times 10^{45}$ \\
\hline
\multicolumn{4}{c}{\textit{Fixed Parameters}} \\
Redshift & $z$ & 0.540 & 1.037 \\
Dissipation distance (cm) & $R_H$ & $2.0 \times 10^{17}$ & $1.0 \times 10^{17}$ \\
Disk Temperature (K) & $T_{\text{disk}}$ & $3.0 \times 10^{4}$ & $3.0 \times 10^{4}$ \\
Inner BLR radius (cm) & $R_{\text{BLR,in}}$ & $3.0 \times 10^{17}$ & $2.0 \times 10^{17}$ \\
Outer BLR radius (cm) & $R_{\text{BLR,out}}$ & $5.0 \times 10^{17}$ & $4.0 \times 10^{17}$ \\
BLR reprocessing & $\tau_{\text{BLR}}$ & 0.1 & 0.1 \\
DT Temperature (K) & $T_{\text{DT}}$ & 1000 & -- \\
DT Radius (cm) & $R_{\text{DT}}$ & $5.0 \times 10^{18}$ & -- \\
DT reprocessing & $\tau_{\text{DT}}$ & 0.1 & -- \\
\hline
\end{tabular}
\par
\vspace{0.1cm}
\raggedright
\footnotesize{\textbf{Note.} --- Quiescent baseline values are derived from fits to archival data for 3C 279 and CTA 102. Parameters listed as ``varied" were free during the minimization process, while ``fixed" parameters were held at fiducial values to constrain the external fields.}
\end{table*}

\bibliography{version1}{}
\bibliographystyle{aasjournalv7}

\end{document}